# EPI-VALID : Validation d'un algorithme d'identification de patients avec épilepsie dans le SNDS à partir des données de la cohorte CONSTANCES.

## Equipe Projet

Pierre-Alain Jachiet, responsable de la mission data, a piloté le projet, géré l'accès aux données de la cohorte Constances, participé à la définition de la méthodologie, participé à l'analyse des résultats, et à la rédaction du rapport.

Adeline Degremont, pharmaco-épidémiologiste à la mission data, a mis en œuvre le projet, défini la méthodologie, interprété et discuté les résultats, rédigé le rapport.

Catherine Bisquay, ingénieure d'étude experte du Système national des données de santé à la mission data, a mis en œuvre le traitement et l'analyse des données, participé à l'analyse des résultats, et à la rédaction du rapport.

Timothée Chehab, ingénieur de données à la mission data, a participé à la gestion de l'accès aux données de la cohorte Constances, et relu le rapport.

## Remerciements

# 1 Synthèse

La mission data de la HAS a mené une étude sur le parcours de soins des patients identifiés comme épileptiques dans le Système national des données de santé (SNDS) en 2018. Cette étude a utilisé **deux algorithmes d'identification des patients avec épilepsie, l'un offrant une estimation haute de la population et l'autre offrant une estimation basse de la population. Ces algorithmes sont** basés sur l'hospitalisation pour épilepsie en établissement de médecin chirurgie obstétrique, l'affection longue durée (ALD) pour épilepsie grave et la délivrance de médicament antiépileptique (AE). **Ils n'ont pas fait l'objet de validation par une source externe**. Pourtant, ce travail de validation apparait nécessaire pour la suite des travaux de la HAS qui consiste à définir et à calculer des indicateurs qualité du parcours à partir des données du SNDS, et qui implique d'identifier une population de patients avec épilepsie. Ainsi, pour étudier les caractéristiques de ces algorithmes SNDS, **nous avons utilisé les données de la cohorte CONSTANCES**, qui sont appariées aux données du SNDS.

CONSTANCES est une cohorte française destinée à fournir des informations descriptives et étiologiques sur la santé publique et l'épidémiologie. Les informations disponibles proviennent de questionnaires remplis par un professionnel de santé et/ou le patient. Dans ces questionnaires, il n'y a pas de questions spécifiques sur l'épilepsie. Ainsi, nous avons seulement pu identifier les patients pour lesquels il y avait une mention d'épilepsie, renseignée en texte libre en réponse à des questions ouvertes du type « Souffrez-vous d'autres pathologies ? ». **Parmi les 156 819 patients présents dans la cohorte CONSTANCES en 2018** (année sur laquelle nous avons appliqué les algorithmes SNDS), **689 patients ont mentionné une épilepsie dans le(s) questionnaire(s) d'inclusion et/ou de suivi, soit une proportion de 0,4%**. Cette proportion est inférieure à la prévalence de l'épilepsie dans la population générale, estimée à 0,6-0,7% chez l'adulte dans la littérature. A partir d'un critère fiable tel que l'ALD pour épilepsie, nous avons pu estimer **qu'au moins un patient sur 3 n'aurait pas mentionné son épilepsie dans la cohorte CONSTANCES**. De plus, la proportion de patients avec une ALD pour épilepsie grave en 2018 est de 0,17% dans la cohorte CONSTANCES alors qu'elle est estimée à 0,32% dans la cohorte témoin de CONSTANCES. Ceci suggère **qu'il y aurait deux fois moins de patients avec épilepsie dans la cohorte CONSTANCES que dans la population générale**.

Connaissant ses deux limites (sous-déclaration et non-représentativité), nous avons **comparé les caractéristiques des deux algorithmes à celles d'un algorithme « contrôle »**. Celui-ci utilise uniquement **l'ALD pour épilepsie grave et l'hospitalisation pour épilepsie** comme critères d'identification dans le SNDS, **puisqu'il s'agit a priori de critères fiables d'identification des patients avec épilepsie**. Les résultats indiquent alors que **la précision des algorithmes SNDS étudiés ici n'est pas très élevée (respectivement 17,8% et 35% en population haute et basse) en comparaison à l'algorithme « contrôle » (58,1%)**. Ces algorithmes en population haute et basse permettent toutefois de **gagner en sensibilité** dans l'identification de patients avec épilepsie par rapport à l'algorithme « contrôle » (+33 points et +14 points respectivement), **sans trop perdre en spécificité**, notamment pour l'algorithme en population basse (-0,3 points versus -1,2 points en population haute).

La perte de précision, que l'on observe en comparaison à l'algorithme « contrôle » est liée au **critère « délivrance de médicaments antiépileptiques »**. A ce propos, l'analyse des patients faux-positifs (FP) nous a permis d'identifier deux pistes d'amélioration. **Premièrement, sélectionner uniquement les patients ayant au moins deux délivrances de médicaments plutôt qu'une seule** permet d'améliorer la précision des algorithmes, notamment pour l'algorithme en population haute (28,6% versus 17,8%) sans altérer (ou peu) leur sensibilité. **Deuxièmement, certains médicaments AE pourraient être exclus des algorithmes car a priori très largement utilisés dans une autre indication**. C'est le cas notamment de la prégabaline et de la gabapentine pour l'algorithme en population haute. Supprimer ces deux médicaments des critères d'inclusion permet d'améliorer nettement la précision de cet algorithme (45,3% versus 17,8%).

**D'autres pistes d'amélioration des algorithmes pourraient également être envisagées** : l'inclusion des séjours hospitaliers pour épilepsie en établissement de soins de suite et réadaptation ou encore l'inclusion des patients sur plusieurs années.

Même s'il faut avoir en tête que les caractéristiques des algorithmes SNDS étudiés ici ne sont pas les valeurs « réelles », les résultats de cette étude éclairent sur les choix à faire quant aux critères d'inclusion à prendre en compte dans le SNDS pour identifier au mieux les patients avec épilepsie.

# Table des matières

## 2 Contexte

Dans le cadre des orientations de la Stratégie de transformation du système de santé (STSS), « ma santé 2022 », la Caisse nationale d'assurance maladie (CNAM) et la Haute autorité de santé (HAS) co-pilotent le projet « inscrire la qualité et la pertinence au cœur des organisations et des pratiques », au sein du chantier 4 de la STSS « Pertinence et qualité ». Le chantier « qualité et pertinence des soins » a pour objectifs la définition d'un parcours, l'élaboration de messages de pertinence et le développement d'indicateurs qualité du parcours. Suite à l'élaboration de la recommandation de bonnes pratiques « Épilepsies : Prise en charge des enfants et des adultes » par la HAS en 2020[1], un travail sur l'amélioration du parcours de soins des patients avec épilepsie est apparu nécessaire en raison de l'absence de filière organisée de soins à l'origine de nombreux dysfonctionnements dans la prise en charge et dans l'accompagnement des patients. L'épilepsie a donc été l'un des 10 parcours retenus. Sur ce projet, la HAS a déjà publié des guides[2] qui décrivent les soins, l'accompagnement et le suivi global de l'adulte et l'enfant avec épilepsie. Ils s'adressent aux professionnels de santé et médicosociaux, aux personnes avec épilepsie et à leur entourage.

Pour accompagner ce projet, et en attendant la définition des indicateurs de qualité du parcours de soins des patients avec épilepsie, la mission data de la HAS a mené une étude sur le parcours de soins des patients identifiés comme épileptiques dans le Système national des données de santé (SNDS) en 2018[3]. Cette étude a utilisé deux algorithmes d'identification des patients avec épilepsie pour le ciblage de la population d'étude basés sur l'hospitalisation pour épilepsie, l'affection longue durée (ALD) pour épilepsie grave et la délivrance de médicament antiépileptique (AE). Les AE peuvent être donnés dans d'autres indications que l'épilepsie. Une première estimation de la population, dite basse, n'a retenu que les médicaments ayant une indication uniquement dans l'épilepsie. Une seconde estimation de la population, dite haute, a utilisé des critères d'exclusion des délivrances de médicaments AE dans un autre contexte que l'épilepsie. Ces critères ont été définis à partir de travaux d'une équipe de pharmaco-épidémiologie de Toulouse (Charlton et al., 2019) et de travaux non publiés de la CNAM sur la prévalence de l'épilepsie en 2014 d'après les données du SNDS (Dantoine et al., 2018), avec l'aide d'un expert neurologue épileptologue. Les algorithmes permettant de sélectionner la population dite haute, et la population dite basse sont présentés dans la Figure 1. Ces algorithmes n'ont pas fait l'objet de validation par une source externe. Or, ce travail de validation apparait nécessaire pour la suite du projet qui consiste à définir et à calculer des indicateurs qualité du parcours à partir des données du SNDS, et qui implique d'identifier une population de patients avec épilepsie.

---

[1] Recommandations disponibles ici : https://www.has-sante.fr/jcms/p_3214468/fr/epilepsies-prise-en-charge-des-enfants-et-des-adultes

[2] Documents sur les guides de parcours disponibles ici : https://www.has-sante.fr/jcms/p_3444925/fr/guides-du-parcours-de-sante-de-l-adulte-et-de-l-enfant-avec-epilepsie

[3] Rapport d'étude disponible ici : https://www.has-sante.fr/upload/docs/application/pdf/2023-06/has-102_rapport_epilepsie_snds_cd_2023-03-23_vd.pdf

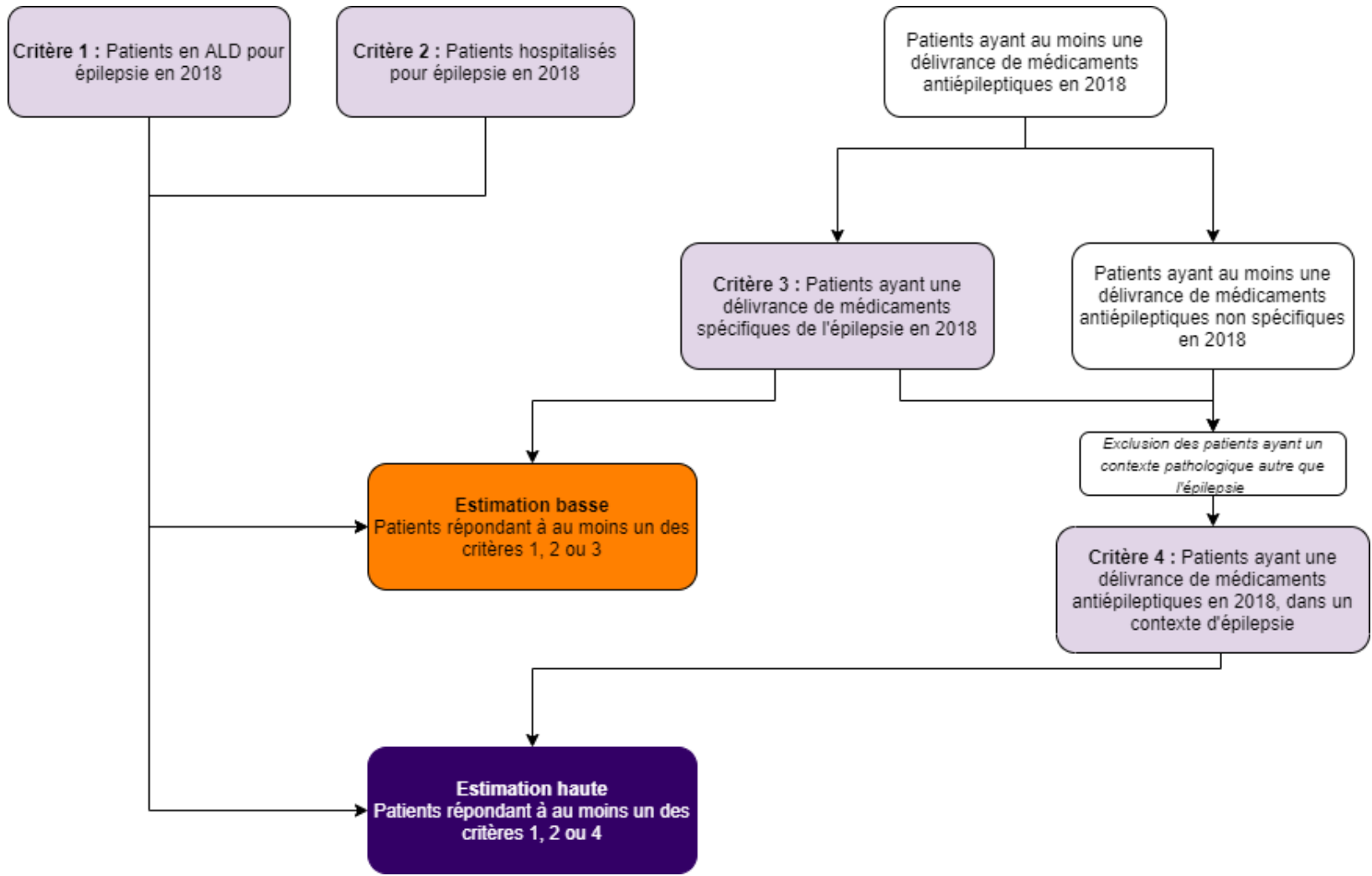

*Figure 1 : Algorithmes d'identification des patients épileptiques dans le Système National des Données de Santé : estimation haute et basse de la population.*

# 3 Objectif

Les algorithmes utilisés pour identifier les patients épileptiques dans l'étude sur le SNDS menée par la mission data n'ayant pas fait l'objet d'une validation, l'objectif du travail EPI-VALID est de **« pré-valider »** (cf. Encadré 1) ces algorithmes en utilisant les données de la cohorte Constances, qui sont appariées aux données du SNDS :

- En identifiant d'une part les patients épileptiques à partir des données du SNDS grâce aux algorithmes SNDS présentés en Figure 1 ;
- En identifiant d'autre part les patients épileptiques à partir des données des questionnaires de la cohorte Constances grâce à la mention d'épilepsie, renseignée (par le médecin et/ou par le patient) en texte libre en réponse à des questions ouvertes des questionnaires d'inclusion et/ou de suivi ;
- En comparant les patients épileptiques identifiés grâce aux algorithmes présentés en Figure 1, à ceux identifiés par la mention d'épilepsie dans les questionnaires d'inclusion et/ou de suivi de la cohorte Constances.

**Encadré 1 : Limites à l'utilisation de la cohorte Constances comme population de référence**

Nous avons utilisé le terme « **pré-valider** » car les données de la cohorte Constances ne sont pas un « gold standard » pour valider le fait que le patient souffre ou non d'épilepsie, et ne peuvent être utilisées pour valider à proprement parler les algorithmes étudiés ici[4].

**D'une part,** parce qu'**on peut s'attendre à une sous déclaration par les patients épileptiques** (et/ou les professionnels de santé)**.** Les données permettant de qualifier le patient comme épileptique sont issues, soit du questionnaire d'inclusion rempli par le médecin, soit du questionnaire de suivi rempli par le patient. Les questionnaires d'inclusion et/ou de suivi n'ont pas de question spécifique sur l'épilepsie. L'épilepsie est renseignée par les patients et/ou les professionnels de santé via des questions ouvertes sur les maladies dont ils souffrent, autres que celles faisant l'objet de questions fermées. Cette sous déclaration est potentiellement renforcée par la stigmatisation associée à cette maladie.

A contrario, on peut penser que la mention d'épilepsie dans les réponses ouvertes est une bonne indication d'une réelle épilepsie. Toutefois, des patients peuvent la renseigner à tort, par exemple dans le cas des crises non-épileptiques psychogènes (CNEP), ou après une première crise d'épilepsie, en attente d'une confirmation diagnostique. Le diagnostic initial peut être posé à tort dans environ 20% des cas, d'après 4 études citées dans l'argumentaire scientifique de la recommandation de bonne pratique de 2020 de la HAS. Le chiffre exact est variable et difficile à estimer (Xu et al., 2016). Ainsi, les patients identifiés comme épileptiques via les questionnaires le sont fort probablement, tandis que la non-identification n'est pas une information fiable.

**D'autre part, se pose la question de la représentativité de la cohorte Constances** vis-à-vis de la population française des patients avec épilepsie. Un biais à l'inclusion peut avoir pour conséquence que la prévalence de l'épilepsie sera différente entre la cohorte Constances et la population générale. Nous avons utilisé l'échantillon témoin pour estimer cette représentativité (cf. 4. Méthode).

Dans la partie discussion, nous avons estimé la valeur de ces deux limites, et discutons leur impact sur le calcul des caractéristiques des algorithmes.

---

[4] Ceci n'est pas vrai pour d'autres pathologies, qui font l'objet de questions fermées. Par exemple, la cohorte Constances a pu être utilisée pour valider des algorithmes d'identification de patients diabétiques : https://www.constances.fr/_assets/_pdf/DIABETE.pdf

# 4 Méthode

## 4.1 Source de données

Les données permettant de « pré-valider » les algorithmes d'identification des patients dans le SNDS proviennent de la cohorte CONSTANCES.

La cohorte CONSTANCES est une cohorte épidémiologique « généraliste » constituée d'un **échantillon représentatif**[5] **d'environ 200 000 adultes, âgés de 18 à 69 ans à l'inclusion**, consultant des centres d'examens de santé (CES) de la sécurité sociale. Cette cohorte comprend des données cliniques recueillies à l'inclusion, et au cours du suivi du patient, appariées aux données du SNDS.

A noter que dans le cadre de cette étude, les données disponibles s'étendaient de 2012 à 2021 pour les questionnaires d'inclusion, et de 2013 à 2020 pour les questionnaires de suivi.

La cohorte Constances contient également les données du SNDS pour un **échantillon témoin d'environ 300 000 personnes** qui permettent d'estimer la représentativité de cette cohorte selon le sujet étudié.

## 4.2 Schéma d'étude

Il s'agit d'une étude de validation d'une méthode d'identification des patients avec épilepsie dans les bases de données médico-administratives, avec comme population d'étude les patients identifiés comme épileptiques d'après les algorithmes SNDS et comme population de référence les patients identifiés comme épileptiques d'après les questionnaires de la cohorte Constances.

### 4.2.1 Population à l'étude

*Population identifiée grâce aux données SNDS*

> *Dans l'étude menée sur le parcours de soins des patients avec épilepsie à partir des données du SNDS par la mission data de la HAS, les algorithmes d'identification des patients épileptiques ont été appliqués uniquement sur l'année 2018. De la même façon, ici, nous avons appliqué les algorithmes sur l'année 2018, en incluant dans notre population source uniquement les patients présents dans la cohorte Constances en 2018 (patients inclus entre 2012 et 2018).*

Tous les patients inclus dans la cohorte Constances avant 2019, et répondant à au moins l'un des critères suivants (cf. Tableau 1) ont été inclus :

---

[5] Les bénéficiaires du régime général sont tirés au sort dans les bases de données de la caisse nationale de l'assurance vieillesse et invités à se rendre dans le CES de leur département affilié à la cohorte Constances pour le recueil des données. Pour plus d'informations sur le tirage au sort des patients et la représentativité de la cohorte d'un point de vue socio-démographique : https://www.constances.fr/base-documentaire/2022/1659537094-resume-du-protocole.pdf

- Présence d'une **ALD pour épilepsie grave** en cours en 2018 ;
- Présence d'une **hospitalisation** (en diagnostic principal [DP], diagnostic relié [DR] ou diagnostic associé [DAS] en établissement de médecine chirurgie obstétrique [MCO]) **pour épilepsie** en 2018 ;
- Présence d'au moins une **délivrance de médicament AE** en 2018,
  - **sauf dans un autre contexte que l'épilepsie**[6], offrant une estimation dite haute de la population épileptique ;
  - **sauf les médicaments ayant d'autres indications** que l'épilepsie dans le cadre de leur autorisation de mise sur le marché (AMM), offrant une estimation dite basse de la population épileptique.

*Tableau 1 : Critères d'inclusion et d'exclusion de la population identifiée dans le SNDS.*

| Critères d'inclusion | Codes/Valeurs |
|---|---|
| ALD pour épilepsie grave en 2018 | Codes de classification internationale des maladies 10ème version (CIM-10) : G40* ou G41*, avec motif d'exonération « ALD » : 41, 43, 45 ou 47 |
| Hospitalisation pour épilepsie en 2018 (mois de sortie) | Codes CIM-10 : G40* ou G41*, en DP/DR/DAS |
| Délivrances de médicaments antiépileptiques en 2018 | Codes anatomique thérapeutique et chimique (ATC) : N03A (antiépileptiques) sauf N03AG02 (DEPAMIDE®) et certains codes identifiant de présentation (CIP) de N03AG01 (DEPAKOTE®), de N05BA09 (clobazam), de N05CD08 (midazolam uniquement sous forme BUCCOLAM®) et de N05BA01 (diazépam uniquement sous forme injectable ou rectale). |

---

[6] Les autres contextes pour lesquels les médicaments AE peuvent être prescrits sont la douleur neuropathique, la migraine, l'anxiété, le trouble bipolaire et la dépendance alcoolique. Ces contextes sont identifiés sur la période 2017 à 2019 grâce à la présence d'une ALD, d'une hospitalisation ou de plusieurs délivrances de médicaments indiqués dans ces contextes (antimigraineux, thymorégulateurs…).

| Critères d'exclusion pour l'estimation haute de la population | Valeurs/Codes |
|---|---|
| ALD pouvant signer une autre indication des traitements AE (années 2017 à 2019) | CIM-10 : G43* (migraine), R52 ou R521 ou R5210 (douleur neuropathique), F10 ou F102* (dépendance alcoolique), F40* ou F41* (anxiété), F30* ou F31* (trouble de l'humeur), avec motif d'exonération « ALD » : 41, 43, 45 ou 47 |
| Hospitalisation pouvant signer une autre indication des traitements AE (années 2017 à 2019) | CIM-10 : G43* (migraine), R5210 (douleur neuropathique), F102* (dépendance alcoolique), F40* ou F41* (anxiété), F30* ou F31* (trouble de l'humeur), en DP/DR/DAS |
| Délivrance de clonazépam en monothérapie | Patients avec au moins une délivrance de clonazépam en 2018 mais pas d'autre AE sur 2017, 2018, 2019 |
| Délivrance de clonazépam en association avec de la prégabaline ou de la gabapentine | Patients avec au moins une délivrance de clonazépam en 2018 et au moins une délivrance de prégabaline ou de gabapentine à la même date |
| Délivrance de prégabaline par un psychiatre | Patients avec au moins une délivrance de prégabaline en 2018 faite par un psychiatre (*Spécialité médicale = 33 ou 75*) |
| Délivrance de carbamazépine ou de lamotrigine par un psychiatre | Patients avec au moins une délivrance de carbamazépine ou de lamotrigine en 2018 faite par un psychiatre (*Spécialité médicale = 33 ou 75*) |
| Délivrances de médicaments pour la dépendance alcoolique | Patients avec au moins 3 délivrances de naltrexone, acamprosate, disulfirame, nalméfène entre 2017 et 2019 |
| Délivrances d'anxiolytiques | Patients avec au moins 3 délivrances de dérivés de la benzodiazépine entre 2017 et 2019 |
| Délivrances de médicaments pour la douleur neuropathique | Patients avec au moins 3 délivrances de capsaicine, duloxétine, amitriptyline, maprotiline, venlafaxine, clomipramine, Imipramine entre 2017 et 2019 |
| Délivrances d'antimigraineux | Patients avec au moins 3 délivrances d'agonistes sélectifs des récepteurs 5HT1, alcaloïdes de l'ergot, pizotifène, erénumab, oxétorone entre 2017 et 2019 |
| Délivrances de thymorégulateurs | Patients avec au moins 3 délivrances de rispéridone, quétiapine, olanzapine, lithium, valpromide entre 2017 et 2019 |
| **Critères d'exclusion pour l'estimation basse de la population** | **Valeurs/codes** |

| Délivrances de médicaments antiépileptiques non spécifiques en 2018 | Cf. Liste des médicaments spécifiques en Annexe 8.1. |
|---|---|

Pour les médicaments, les tables SNDS requêtées pour l'identification des patients ont été :

- ER_PHA_F (pharmacie de ville) et ER_UCD_F (rétrocession) pour le datamart de consommations inter-régimes (DCIR) ;

- les tables T_MCOaaMED et T_MCOaaMEDATU qui concernent les médicaments délivrés en sus du groupe homogène de séjour (GHS) ou faisant l'objet d'un dispositif d'autorisation temporaire d'utilisation (ATU) pour le Programme de médicalisation du système d'information (PMSI).

A noter qu'aucune délivrance n'a été retrouvée dans les tables du PMSI.

### 4.2.2   Population de référence

*Population identifiée grâce aux données recueillies dans la cohorte Constances*

> *Pour rappel, dans l'étude menée sur le parcours de soins des patients avec épilepsie à partir des données du SNDS, les algorithmes d'identification des patients épileptiques ont été appliqués uniquement sur l'année 2018 dans l'objectif d'identifier les patients épileptiques en 2018. Notre population source comprend donc tous les patients présents dans la cohorte Constances en 2018, c'est-à-dire tous les patients inclus entre 2012 et 2018.*

Tous les patients pour lesquels on retrouve une mention d'une épilepsie dans leur questionnaire d'inclusion et/ou de suivi pour toutes les années antérieures à 2019 ont été inclus.

Pour rappel, l'objectif était d'identifier tous les patients épileptiques en 2018, pour pouvoir correspondre à l'algorithme SNDS, appliqué sur l'année 2018. Or, dans les questionnaires Constances, il n'est pas possible de connaître la date de début de l'épilepsie. Parmi les patients pour lesquels on retrouve une mention d'épilepsie, plusieurs situations différentes peuvent donc être retrouvées :

- mention avant 2018 ;
- mention uniquement après 2018, pour une maladie déjà présente en 2018 ;
- mention uniquement après 2018 pour une maladie qui débute après 2018.

Ainsi, pour éviter d'inclure à tort des patients non épileptiques en 2018, nous avons fait le choix de ne pas inclure les deux dernières situations (mention après 2018) et seules les données des questionnaires d'inclusion et/ou de suivi recueillies entre 2012 et 2018 ont été utilisées pour l'identification des patients épileptiques.

L'information sur la présence d'une épilepsie est non structurée dans Constances : elle est renseignée dans des champs en texte libre. Les mots clés suivants ont donc été recherchés : épilepsie, absence, comitialité, merrf, hippocampectomie, petit mal et grand mal (hors accent, hors casse). Nous avons choisi de ne pas prendre en compte le mot clé convulsion, qui risquait d'inclure à tort des patients non épileptiques. Une vérification manuelle a ensuite été menée pour intégrer toutes les orthographes possibles de ces mots clés (par exemple : comotialite, conitialite, comiciate, epilepcie, epilespie, …). A noter que six cas ont été classés en suspicion car considérés comme douteux (par exemple : crise probable d'épilepsie, cf. Annexe 8.2), mais ont tout de même été inclus dans la population identifiée comme épileptique.

Les données des questionnaires permettant d'identifier les patients avec épilepsie étaient :

Questionnaire Constances d'Inclusion (formulaire de l'année 2022) :

- 5_Questionnaire médical ; Partie : Antécédents médicaux personnels ; Questions n°1 à 11 concernant les autres affections (cardiovasculaires, respiratoires, digestives, uro-génitales, rénales, neurologiques et psychiques, ostéoarticulaires, endocriniennes, cancer, autres)

> *Nous avons fait le choix d'inclure toutes ces questions, même si, a priori, l'épilepsie devrait être renseignée dans le champ « autres affections neurologiques et psychiques ».*

Questionnaire Constances de Suivi (formulaire de l'année 2022) :

- 2_Santé I, Echelles de santé ; Partie : Liste des pathologies ; Sous-partie : Autres ; Question : Autres problèmes de santé, y compris psychique.

## 4.3 Variables

Conformément aux paragraphes ci-dessus, les données recueillies pour la sélection des patients étaient :

- Les dates de début et de fin de l'ALD, ainsi que les motifs ;
- Les dates de séjours hospitaliers et leurs motifs (en DP/DR/DAS) ;
- Les dates de délivrances de médicaments AE et les molécules délivrées ;
- La mention d'une épilepsie dans les questionnaires de suivi ou d'inclusion.

## 4.4 Analyse statistique

Le logiciel de traitement et d'analyse des données utilisé est SAS®.

Afin de comparer nos deux populations, l'une issue de l'algorithme dans le SNDS (population à l'étude), l'autre issue des données des questionnaires (population de référence), nous avons calculé pour chaque algorithme (en population haute ou basse) :

- le nombre de patients vrais positifs (VP) correspondant au nombre de patients avec une mention d'épilepsie dans les questionnaires Constances retrouvés d'après l'algorithme SNDS ;

- le nombre de patients faux négatifs (FN), correspondant au nombre de patients avec une mention d'épilepsie dans les questionnaires Constances non retrouvés d'après l'algorithme SNDS ;

- le nombre de patients vrais négatifs (VN), correspondant au nombre de patients sans mention d'épilepsie dans les questionnaires Constances non retrouvés d'après l'algorithme SNDS ;

- le nombre de patients faux positifs (FP), correspondant au nombre de patients identifiés comme épileptiques d'après l'algorithme SNDS sans mention d'épilepsie dans les questionnaires Constances.

A partir de ces nombres, nous avons calculé les caractéristiques de chaque algorithme :

- la valeur prédictive positive (VPP) ou précision de l'algorithme par rapport aux questionnaires qui correspond à la proportion de patients épileptiques parmi les patients identifiés via l'algorithme (VP/VP+FP) ;
- la valeur prédictive négative (VPN), qui correspond à la proportion de patients non épileptiques parmi les patients non identifiés comme tels via l'algorithme (VN/VN+FN) ;
- la spécificité qui rapporte la capacité à ne détecter que les malades (VN/VN+FP) ;
- la sensibilité qui rapporte la capacité à détecter un maximum de malades (VP/VP+FN).

---

**Encadré 2 : Comparaison à un algorithme dit « contrôle »**

Les caractéristiques de chaque algorithme ont été calculées tout en sachant qu'elles ne correspondent pas aux caractéristiques réelles en raison de la source de données utilisée (cf. Encadré 1). L'interprétation de ces caractéristiques devra se faire avec prudence. Pour évaluer les caractéristiques de chaque algorithme, nous avons choisi de les comparer à un algorithme dit « contrôle » ne prenant pas en compte le critère délivrance de médicaments antiépileptiques. Pour étudier les caractéristiques de cet algorithme dit « contrôle », la population à l'étude correspondait à tous les patients présents dans la cohorte Constances avant 2019, qui présentaient au moins l'un des critères suivants :

- Présence d'une **ALD pour épilepsie grave** en cours en 2018 ;
- Présence d'une **hospitalisation** (en DP, DR ou DAS en établissement de MCO) **pour épilepsie** en 2018.

Ces critères d'ALD ou d'hospitalisation pour épilepsie ont été choisis car ils sont « spécifiques » de l'épilepsie ; ce sont d'ailleurs les critères choisis par la CNAM pour la cartographie des pathologies et des dépenses associées jusqu'en 2022[7] (appliqués sur plusieurs années, à défaut d'une seule ici).

De la même façon que pour les algorithmes en population haute et basse, cette population a été comparée à la population de référence définie en 3.2.2.

---

[7] L'Assurance Maladie met à disposition du grand public un ensemble de données sur une cinquantaine de pathologies, traitements chroniques et épisodes de soins https://data.ameli.fr/pages/data-pathologies/

# 5 Résultats

## 5.1 Concernant l'algorithme en population haute

### 5.1.1 Nombre de patients et tableau croisé

N = 156 819 patients présents dans la cohorte CONSTANCES en 2018.

**Les patients CONSTANCES +** sont les patients ayant renseigné une **épilepsie dans le(s) questionnaire(s)** d'inclusion et/ou de suivi ; **ils sont au nombre de 689**. Inversement, les patients CONSTANCES - sont ceux n'ayant pas fait mention d'une épilepsie (cf. Tableau 2).

**Les patients SNDS +** sont ceux identifiés comme épileptiques selon l'algorithme SNDS ; **ils sont au nombre de 2 453**. Inversement, les patients SNDS- sont ceux non identifiés comme épileptiques selon l'algorithme (cf. Tableau 2).

*Tableau 2 : Tableau croisé des patients dans le SNDS et des patients dans CONSTANCES avec épilepsie – Algorithme en population haute.*

|  | CONSTANCES + | CONSTANCES - | Total |
|---|---|---|---|
| **SNDS +** | 437 = VP | 2 016 = FP | **2 453** |
| **SNDS -** | 252 = FN | 154 114 = VN | 154 366 |
| Total | **689** | 156 130 | 156 819 |

### 5.1.2 Caractéristiques de l'algorithme

La sensibilité est égale à : $\frac{VP}{VP+FN} = \frac{437}{689} = \mathbf{63,4}\%$

La spécificité est égale à : $\frac{VN}{VN+FP} = \frac{154\,114}{156\,130} = \mathbf{98,7}\%$

La VPP est égale à : $\frac{VP}{VP+FP} = \frac{437}{2\,453} = \mathbf{17,8}\%$

La VPN est égale à : $\frac{VN}{VN+FN} = \frac{154\,114}{154\,366} = \mathbf{99,8}\%$

### 5.1.3 Analyse des critères d'inclusion

#### 5.1.3.1 ALD et hospitalisation

Parmi les patients en **ALD pour épilepsie grave** (N=263), 185 **(70%) ont déclaré une épilepsie** dans Constances. Parmi les patients avec une **hospitalisation pour épilepsie (sans ALD)** (N=97), 24 **(25%) ont déclaré une épilepsie** dans Constances. La Figure 2 offre une autre représentation de ces données.

La proportion de patients VP avec une ALD pour épilepsie grave est de 42%, et celle de patients FP avec une ALD est de 4%. La proportion de patients VP avec une hospitalisation pour épilepsie (avec ou sans ALD) est de 12% et celle de patients FP avec une hospitalisation est de 4%. Ces proportions sont représentées selon un diagramme de Venn en Figure 3. L'ensemble des données est disponible en Annexe 8.3.

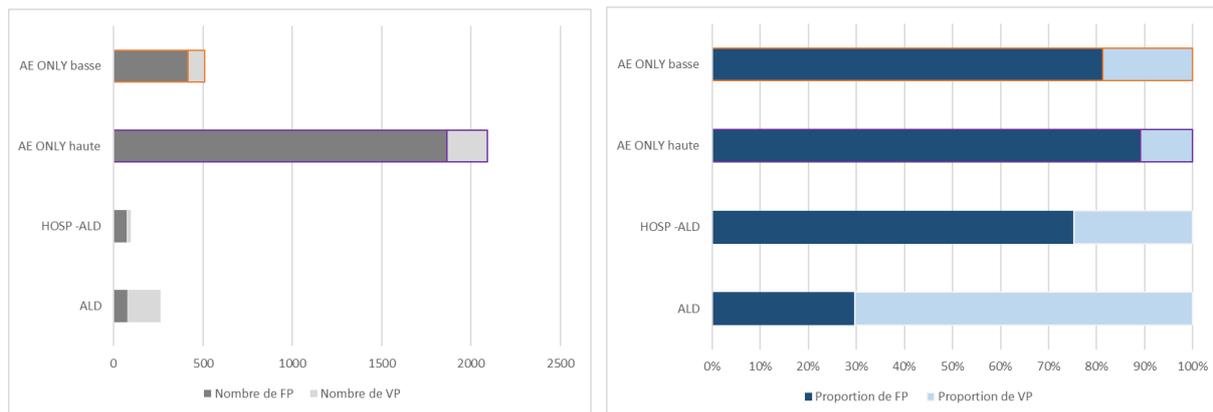

*Figure 2 : Nombre (à gauche) et proportion (à droite) de patients VP et FP, selon le critère d'identification : ALD, hospitalisation (sans ALD), délivrance de médicament AE (sans ALD ni hospitalisation) d'après l'algorithme en population haute et basse.*

### 5.1.3.2 Délivrance de médicaments AE

Parmi les patients avec une **délivrance de médicaments AE (sans ALD ni hospitalisation)** (N=2093), 228 **(11%) ont déclaré une épilepsie dans Constances** (cf. Figure 2). Ces patients représentent la **moitié des patients VP**. Parmi eux **97,4% (N=222) ont eu au moins deux délivrances d'AE** à au moins 30 jours d'intervalle. **Parmi les FP, 92,5%** (N=1865) sont concernés par une délivrance de médicament AE, sans ALD, ni hospitalisation, et parmi eux **la moitié (N=925, 49,6%) a eu au moins deux délivrances d'AE** à au moins 30 jours d'intervalle.

### 5.1.4 Analyse des FN

La population compte 252 FN, soit **37% de la population CONSTANCES + non identifiée via l'algorithme SNDS**. L'analyse de ces FN révèle que :

- 18% (N=45) avaient une délivrance d'antiépileptique en 2018 mais présentaient des critères d'exclusion (autre contexte pathologique)[8] ;
- 6% (N=16) avaient une délivrance d'antiépileptique ou une ALD pour épilepsie grave ou une hospitalisation pour épilepsie en 2017 ou 2019 mais pas en 2018 ;
- 76% (N=191) n'avaient ni délivrance d'antiépileptique, ni ALD pour épilepsie, ni hospitalisation pour épilepsie sur la période 2017 à 2019, et parmi eux :
  - ○ 7 patients avaient eu au moins un électroencéphalogramme (EEG) entre 2017 et 2019, soit environ 4% de ces patients ;
  - ○ 3 patients avaient été classés en suspicion d'épilepsie, car ils n'avaient renseigné dans les questionnaires qu'une crise isolée ou probable d'épilepsie (cf. Annexe 8.2), soit moins de 2% de ces patients ;
  - ○ 2 patients avaient une ALD pour AVC entre 2017 et 2019, soit 1% de ces patients ;

---

[8] Sur l'ensemble des patients ayant une délivrance d'AE en 2018 (N=4234), 1933 patients ont été exclus car ils présentaient des critères d'exclusion (autre contexte pathologique) (46%).

o Pour les patients restants, nous n'avons retrouvé aucun contexte clinique entre 2017 et 2019, à savoir ni hospitalisation (en MCO), ni ALD, ni événement traceur pouvant indiquer une épilepsie.

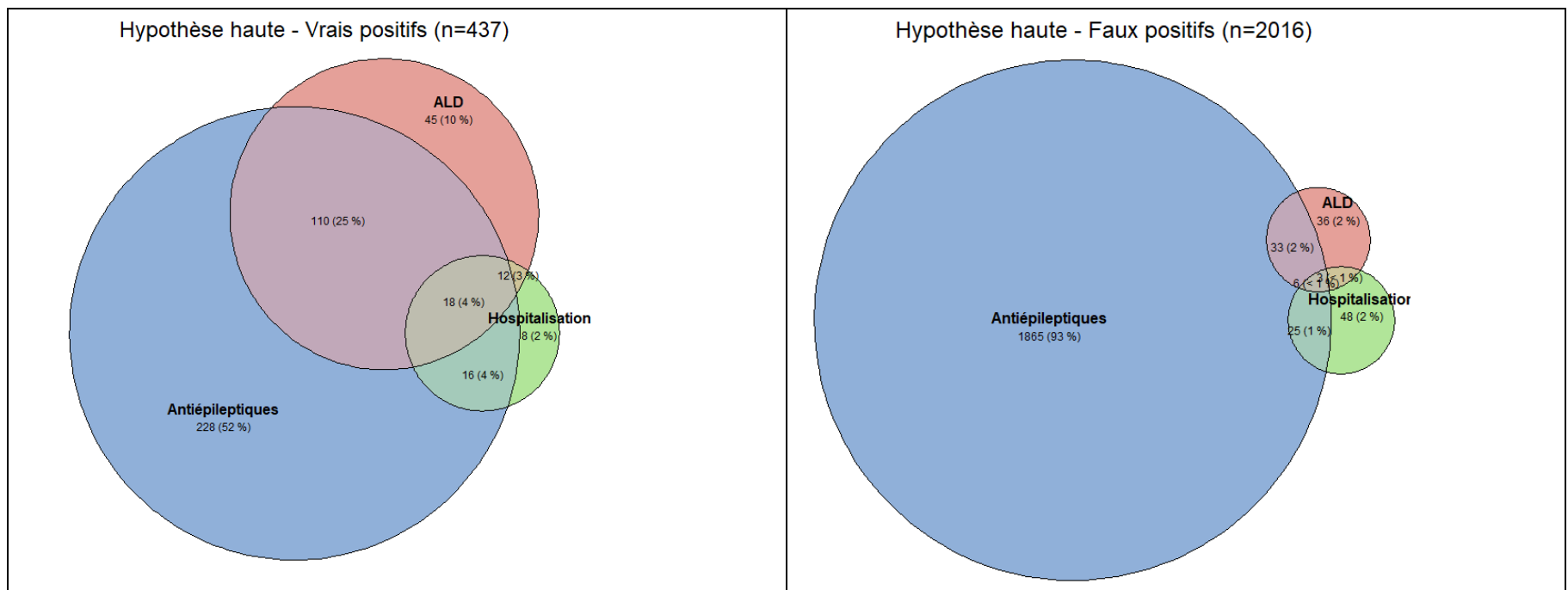

*Figure 3 : Diagramme de Venn représentant les modalités d'inclusion des patients VP (à gauche) et FP (à droite) – Algorithme en population haute*

### 5.1.5  Analyse des FP

Une analyse individuelle des patients FP n'était pas possible au regard du nombre de patients concernés (N=2016). Notons toutefois que 92,5% des FP ont au moins une délivrance d'AE, sans ALD, ni hospitalisation. Nous avons analysé les molécules délivrées parmi ces patients (N= 1865) en comparaison aux molécules délivrées parmi les patients VP identifiés de la même façon (N=228). Cette analyse nous indique qu'**un très grand nombre de patients FP sont identifiés par une délivrance de prégabaline ou de gabapentine (64% et 18% des patients FP respectivement).** Les résultats sont détaillés dans l'Annexe 8.4. Par ailleurs, les patients ayant une délivrance de prégabaline ou de gabapentine sont majoritairement des patients FP (cf. Figure 4) (99,7% et 98,6% respectivement). Enfin, il est à noter que les patients ayant une délivrance de diazépam ou de clonazépam sont tous des patients FP, même si leur proportion sur le total des patients FP est faible (cf. Annexe 8.4).

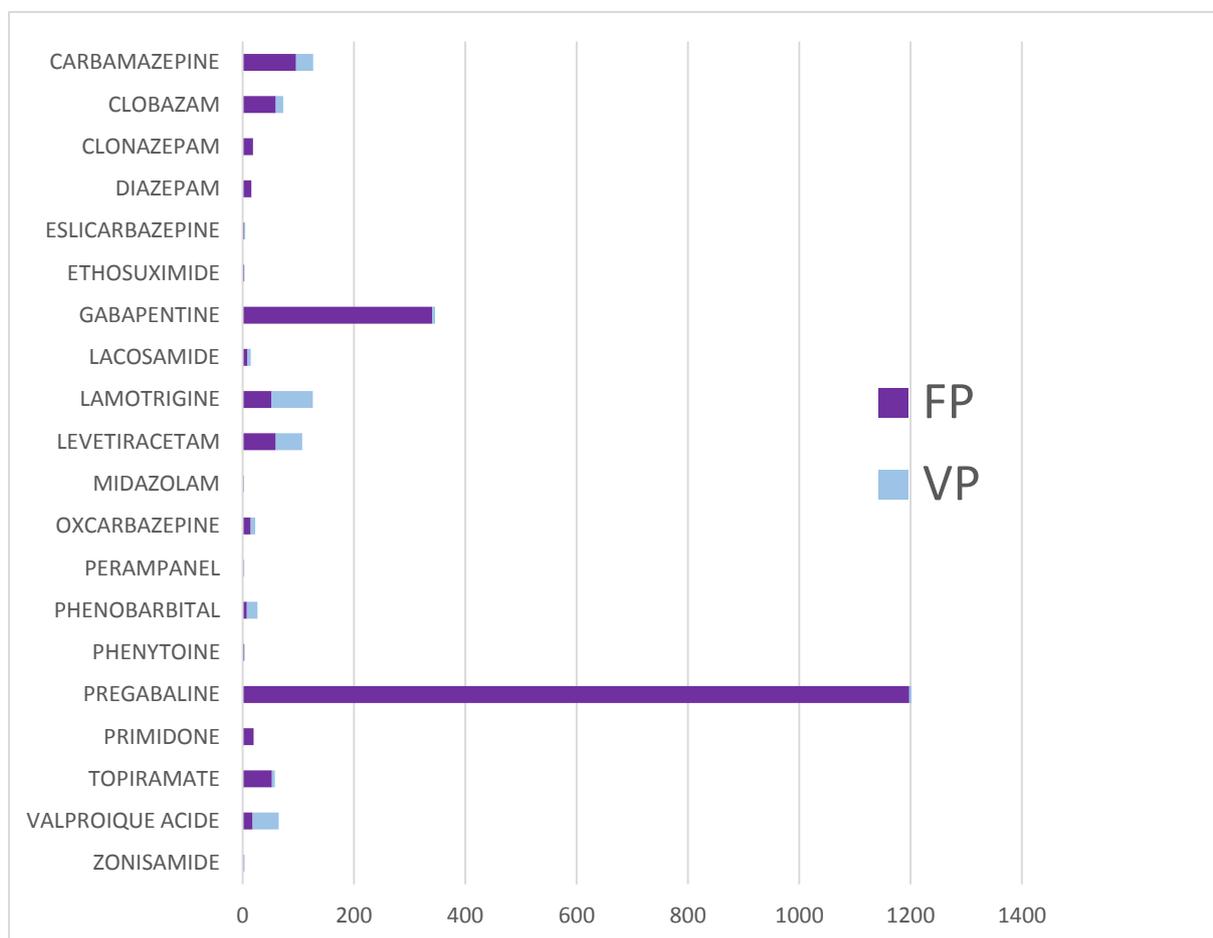

*Figure 4 : Répartition du nombre de patients VP et FP par AE dans la population haute ; patients identifiés uniquement par le critère médicament AE (à savoir sans ALD ni hospitalisation pour épilepsie). A noter qu'un même patient peut être comptabilisé pour deux médicaments différents*

## 5.2  Concernant l'algorithme en population basse

### 5.2.1  Nombre de patients et tableau croisé

Pour rappel, 156 819 patients sont présents dans la cohorte CONSTANCES en 2018, dont **689 ont déclaré une épilepsie (CONSTANCES +**).

Les patients **SNDS+** sont ceux identifiés comme épileptiques selon l'algorithme SNDS en population dite basse ; **ils sont au nombre de 871**.

*Tableau 3 : Tableau croisé des patients dans le SNDS et des patients dans CONSTANCES avec épilepsie – Algorithme en population basse.*

|  | **CONSTANCES +** | **CONSTANCES -** | Total |
|---|---|---|---|
| **SNDS +** | 305 = VP | 566 = FP | **871** |
| **SNDS -** | 384 = FN | 155 564 = VN | 155 948 |
| Total | **689** | 156 130 | 156 819 |

### 5.2.2   Caractéristiques de l'algorithme

La sensibilité est égale à : $\frac{VP}{VP+FN} = \frac{305}{689} = \mathbf{44,3}\%$

La spécificité est égale à : $\frac{VN}{VN+FP} = \frac{155\,564}{156\,130} = \mathbf{99,6}\%$

La VPP est égale à : $\frac{VP}{VP+FP} = \frac{305}{871} = \mathbf{35}\%$

La VPN est égale à : $\frac{VN}{VN+FN} = \frac{155\,564}{155\,948} = \mathbf{99,8}\%$

### 5.2.3   Analyse des critères d'inclusion

#### 5.2.3.1   ALD et hospitalisation

Pour rappel, parmi les patients en ALD pour épilepsie grave (N=263), 185 (70%) ont déclaré une épilepsie dans Constances et parmi les patients avec une hospitalisation pour épilepsie (sans ALD) (N=97), 24 (25%) ont déclaré une épilepsie dans Constances (cf. Figure 2).

La proportion de patients VP avec une ALD est de 61%, et celle de patients FP avec une ALD est de 14% (cf. Diagrammes de Venn Figure 6 et Tableaux en Annexe 0). La proportion de patients VP avec une hospitalisation pour épilepsie (avec ou sans ALD) est de 18% et celle de patients FP avec une hospitalisation est de 14%.

#### 5.2.3.2   Délivrance de médicaments AE

Parmi les patients identifiés par une **délivrance de médicament AE, sans ALD, ni hospitalisation** (N= 511), 96 **(19%) ont déclaré une épilepsie dans Constances** (cf. Figure 2). Ces patients représentent **un tiers des patients VP** ; parmi eux **95% (N=91) ont eu au moins deux délivrances d'AE** à au moins 30 jours d'intervalle. **Parmi les FP, 73%** (N=415) sont concernés par une délivrance de médicament AE, sans ALD, ni hospitalisation, et parmi eux **64% (N=267) ont eu au moins deux délivrances d'AE** à au moins 30 jours d'intervalle

### 5.2.4 Analyse des FN

Cette analyse n'a pas été réalisée pour cet algorithme. Elle n'est pas pertinente en population basse, puisque de fait la population est dite basse car elle exclut de nombreux patients épileptiques traités par un AE non spécifique de l'épilepsie.

Pour rappel, l'analyse des FN a été réalisée pour l'algorithme en population dite haute, afin notamment d'identifier le nombre de patients exclus à tort par application des critères d'exclusion.

### 5.2.5 Analyse des FP

La population compte 566 patients FP. Parmi ces patients, rappelons que 14% ont une ALD et 14% ont une hospitalisation pour épilepsie. Notons que 73% des patients FP sont identifiés uniquement par une délivrance d'AE, sans ALD ni hospitalisation pour épilepsie. Pour ces 415 patients, nous avons analysé la répartition des molécules délivrées, en comparaison aux molécules délivrées parmi les patients VP identifiés de la même façon (N=96) Cette analyse nous indique qu'un très **grand nombre de patients FP sont identifiés par une délivrance de clonazépam (41%)**. Les résultats sont détaillés dans l'Annexe 8.4. Par ailleurs, les patients ayant une délivrance de clonazépam sont majoritairement des patients FP (98,8%) (cf. Figure 5). Notons qu'il en est de même pour les patients ayant une délivrance de diazépam et de primidone qui sont tous des patients FP, ou encore les patients ayant une délivrance d'oxcarbazépine, qui se retrouvent très majoritairement dans le groupe des patients FP (83%) (cf. Figure 5), bien que ces patients ne représentent qu'une faible proportion de l'ensemble des patients FP (identifiés uniquement par une délivrance d'AE, sans ALD ni hospitalisation pour épilepsie, pour rappel) (cf. Annexe 8.4).

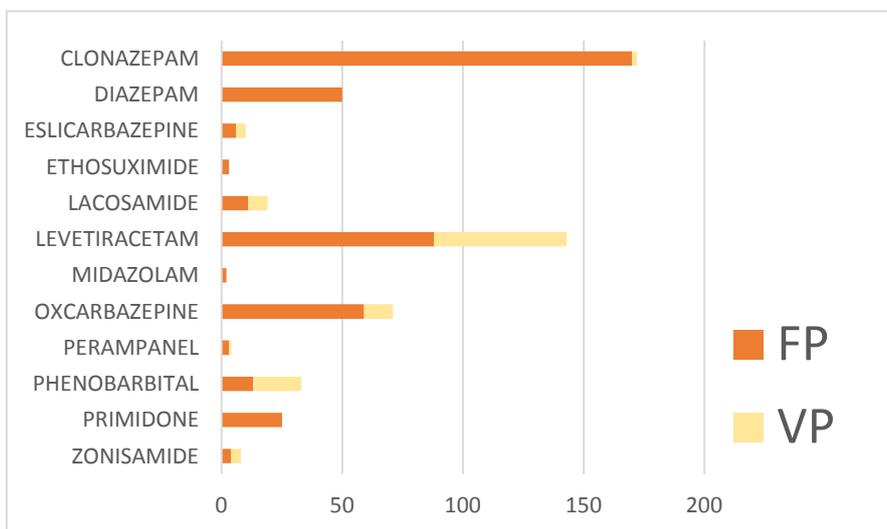

*Figure 5 : Répartition du nombre de patients VP et FP par AE dans la population basse ; patients identifiés uniquement par le critère médicament AE (à savoir sans ALD ni hospitalisation pour épilepsie). A noter qu'un même patient peut être comptabilisé pour deux médicaments différents.*

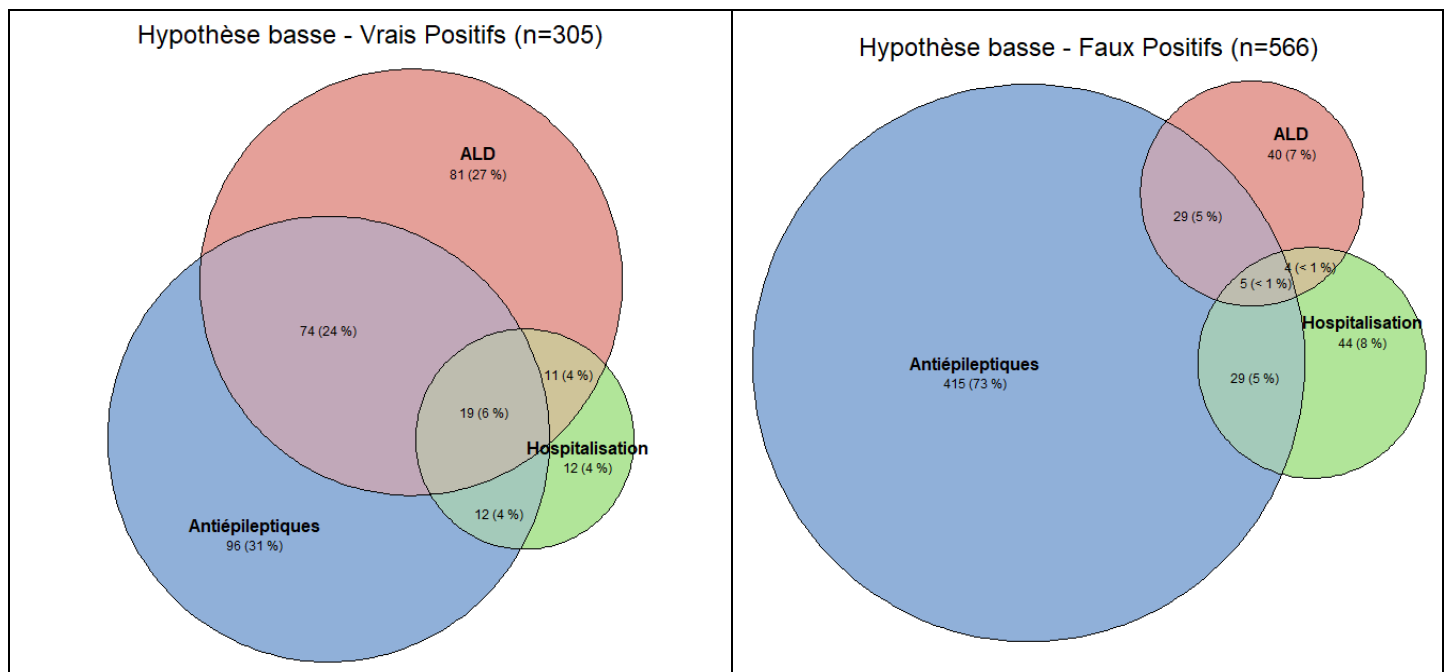

*Figure 6 : Diagramme de Venn représentant les modalités d'inclusion des patients VP (à gauche) et FP (à droite) – Algorithme en population haute*

## 5.3   Concernant l'algorithme « de contrôle »

### 5.3.1   Nombre de patients et tableau croisé

Pour rappel, 156 819 patients sont présents dans la cohorte CONSTANCES en 2018, dont **689 ont déclaré une épilepsie (CONSTANCES +**).

Les patients **SNDS+** sont ceux identifiés comme épileptiques selon l'algorithme SNDS « de contrôle » ; **ils sont au nombre de 360**.

*Tableau 4 : Tableau croisé des patients dans le SNDS et des patients dans CONSTANCES avec épilepsie – Algorithme « de contrôle ».*

|         | CONSTANCES +   | CONSTANCES -       | Total   |
|---------|----------------|--------------------|---------|
| **SNDS +** | 209 = VP    | 151 = FP           | **360** |
| **SNDS -** | 480 = FN    | 155 979 = VN       | 156 459 |
| Total   | **689**        | 156 130            | 156 819 |

### 5.3.2   Caractéristiques de l'algorithme

La sensibilité est égale à : $\frac{VP}{VP+FN} = \frac{209}{689} = \mathbf{33,3}\%$

La spécificité est égale à : $\frac{VN}{VN+FP} = \frac{155\,979}{156\,130} = \mathbf{99,9}\%$

La VPP est égale à : $\frac{VP}{VP+FP} = \frac{209}{360} = \mathbf{58,1}\%$

La VPN est égale à : $\frac{VN}{VN+FN} = \frac{155\,979}{156\,459} = \mathbf{99,7}\%$

### 5.3.3   Analyse des critères d'inclusion

Pour rappel, parmi les patients en ALD pour épilepsie grave (N=263), 185 (70%) ont déclaré une épilepsie dans Constances et parmi les patients avec une hospitalisation pour épilepsie (sans ALD) (N=97), 24 (25%) ont déclaré une épilepsie dans Constances (cf. Figure 2).

La proportion de patients VP avec une ALD pour épilepsie grave est de 89%, et celle de patients FP avec une ALD est de 52%. La proportion de patients VP avec une hospitalisation pour épilepsie (avec ou sans ALD) est de 26% et celle de patients FP avec une hospitalisation est de 54%. Ces proportions sont représentées selon un diagramme de Venn en Figure 7

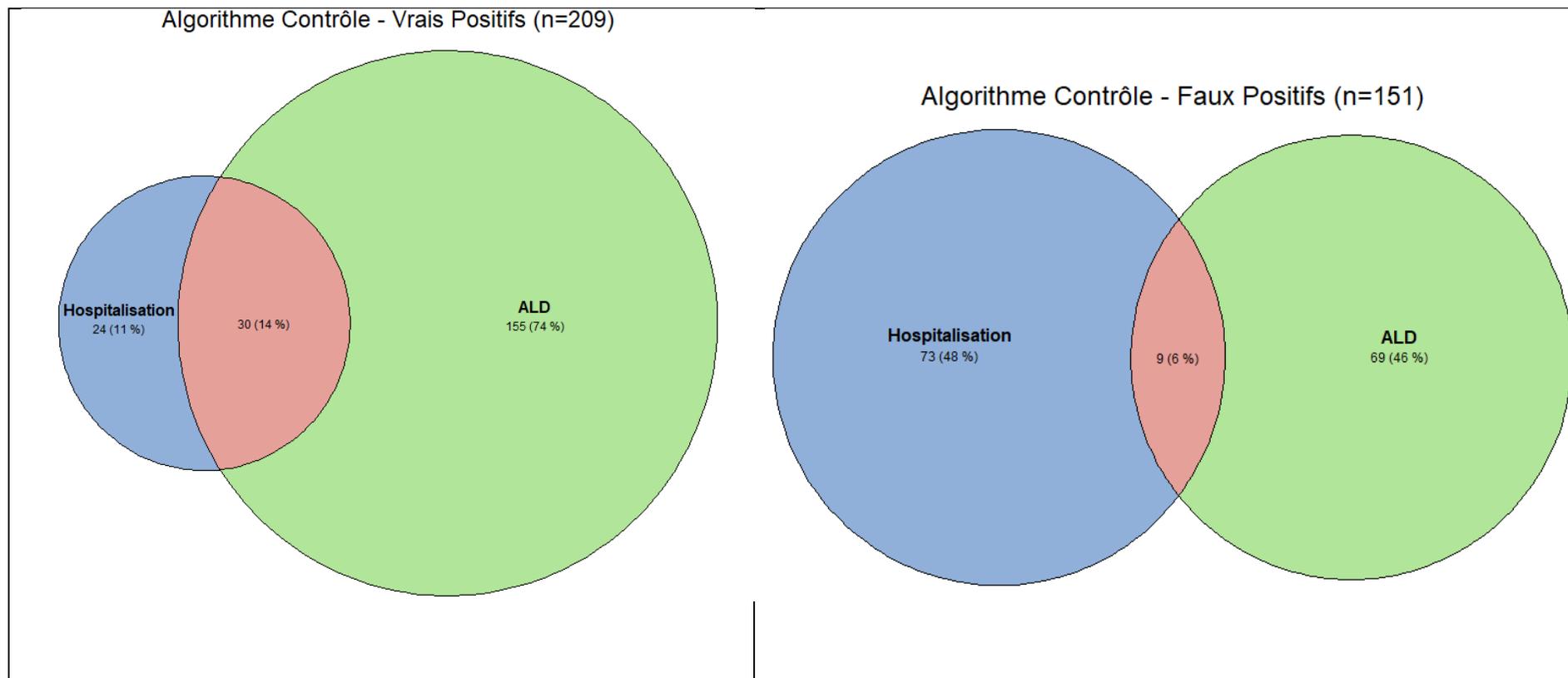

*Figure 7 : Diagramme de Venn représentant les modalités d'inclusion des patients VP (à gauche) et FP (à droite) – Algorithme « contrôle »*

# 6 Discussion

## 6.1 Nombre de patients avec épilepsie

Parmi les 156 819 patients présents dans la cohorte CONSTANCES en 2018, **689 patients** ont mentionné une **épilepsie dans le(s) questionnaire(s)** d'inclusion et/ou de suivi, soit une proportion de 0,4%.

Ces patients sont identifiés comme épileptiques majoritairement à l'inclusion (N=497, 72,7%).

En comparaison aux données de la littérature sur la prévalence de l'épilepsie, cette proportion de patients avec épilepsie apparaît faible. En effet, d'après les données de la littérature (Forsgren et al., 2005), la prévalence de l'épilepsie est estimée à 0,6% chez l'adulte de 20 à 64 ans et 0,7% au-delà de 65 ans. Cette différence s'explique par les limites de la cohorte Constances évoquées dans l'Encadré 1 : l'absence de représentativité et la sous-déclaration, que l'on détaille ci-dessous.

*Absence de représentativité*

Pour rappel, pour estimer la représentativité de la cohorte Constances par rapport à la population française des patients avec épilepsie, nous avons utilisé l'échantillon témoin. Dans cet échantillon témoin, la proportion de patients avec une ALD pour épilepsie grave en 2018 est de 0,32% et la proportion de patients avec une hospitalisation pour épilepsie en 2018 est de 0,18% (cf. Tableau 5 : Estimation de la représentativité– Calculs sur l'année 2018.). Dans la cohorte Constances, on estime respectivement à 0,17% et 0,09% ces mêmes proportions. **Ceci suggère qu'il y a deux fois moins de patients avec épilepsie dans la cohorte Constances que dans la population générale**, même s'il n'est pas possible d'extrapoler ces calculs à l'ensemble de la population avec épilepsie car tous les patients avec épilepsie ne sont pas concernés par une hospitalisation pour épilepsie dans l'année, ni par une ALD pour épilepsie grave, et que le biais de sélection dans Constances peut dépendre de la gravité de l'épilepsie.

*Tableau 5 : Estimation de la représentativité– Calculs sur l'année 2018.*

|  | Cohorte témoin | | Cohorte Constances | |
|---|---|---|---|---|
| **TOTAL** | **304 780** | | **156 819** | |
| ALD pour épilepsie grave | 989 | 0,32% | 263 | 0,17% |
| Hospitalisation pour épilepsie | 536 | 0,18% | 136 | 0,09% |

*Sous-déclaration*

Pour estimer la sous-déclaration, nous avons recensé le nombre de patients en ALD pour épilepsie grave qui n'ont pas déclaré leur épilepsie dans la cohorte Constances, l'ALD apparaissant comme un critère fiable d'identification des patients avec épilepsie. Trente pour cent (N=78/263) des patients en ALD pour épilepsie grave ne l'ont pas déclaré dans la cohorte Constances (cf. Tableau 6). Ceci suggère qu'**un patient sur trois n'aurait pas mentionné son épilepsie dans la cohorte Constances.** De plus, la proportion de patients ayant fait mention d'une épilepsie dans Constances est plus importante parmi les patients en ALD pour épilepsie grave (70%) que parmi les patients hospitalisés (25%) ou encore les patients ayant une délivrance d'AE (11% pour l'algorithme en population haute, et 19% pour l'algorithme en population basse). Ces résultats indiquent que **les patients atteints d'épilepsie plus sévère ont tendance à davantage déclarer leur maladie**. Au total, il existe un facteur de sous-déclaration des patients avec épilepsie dans la cohorte Constances **d'au moins** un patient sur trois.

*Tableau 6 : Estimation de la sous-déclaration– Calculs sur l'année 2018*

|  | Cohorte Constances | Constances + | |
|---|---|---|---|
| **TOTAL** | **156 819** | **689** | |
| ALD pour épilepsie grave | 263 | 185 | 70% |
| Hospitalisation pour épilepsie | 136 | 54 | 40% |
| Hospitalisation pour épilepsie sans ALD | 97 | 24 | 25% |
| ALD et/ou Hospitalisation pour épilepsie | 360 | 209 | 58% |

**En conclusion**, on sait que dans la cohorte Constances, il y a moins d'épileptiques que dans la population générale, et que tous ne le déclarent pas. A contrario, concernant les algorithmes SNDS, on retrouve **2 à 3 fois plus de patients d'après l'algorithme offrant une estimation dite haute de la population, et une proportion similaire de patients d'après l'algorithme offrant une estimation dite basse de la population**, en comparaison aux données de la littérature**.** En effet, on retrouve 2 453 patients avec une épilepsie sur 156 819 participants, soit une proportion de 1,5% d'après l'algorithme offrant une estimation haute de la population, et 871 patients, soit une proportion de 0,56% d'après l'algorithme offrant une estimation basse de la population.

.

## 6.2 Caractéristiques des algorithmes et critères d'inclusion

### 6.2.1 *VPP et VPN*

La probabilité que la maladie soit présente lorsque l'algorithme la détecte (VPP) n'est que de 17,8% en population dite haute, et 35% en population dite basse.

La probabilité que la maladie ne soit pas présente lorsque l'algorithme ne la détecte pas (VPN) est de 99,8% en population dite haute et basse.

La VPP apparaît basse et reflète deux éléments :

- la sous-déclaration des patients avec épilepsie dans Constances (nombre de FP élevé car non recensés dans Constances) ;
- et la présence de patients réellement non épileptiques inclus via les algorithmes SNDS (nombre réel de FP élevé).

Il faut donc avoir en tête que la précision « vraie » des algorithmes est supérieure à ces estimations, sans que nous puissions précisément la déterminer. En l'absence de sous-déclaration dans Constances, la précision d'un algorithme de ciblage tel que l'algorithme dit « de contrôle » devrait être proche de 100%. Ici, donc en présence de sous-déclaration, on retrouve une précision de 58,1%. Ainsi, la précision « vraie » d'un algorithme incluant d'autres critères que l'ALD et l'hospitalisation, à savoir les médicaments, devrait augmenter d'un facteur multiplicatif similaire (x1,7 environ = 100/58).

Par ailleurs, nous avons montré que la cohorte Constances n'était pas représentative de la population générale, et qu'il y avait a priori deux fois moins de patients avec épilepsie. Or, une multiplication par deux de la prévalence a un impact important sur la précision (VPP) d'un algorithme, lorsque la prévalence de la maladie est faible, comme c'est le cas pour l'épilepsie. Ce facteur de sous-estimation de la précision (liée à la faible prévalence de l'épilepsie) est indépendant et s'ajoute à celui de la sous-déclaration.

### 6.2.2 *Sensibilité et spécificité*

**L'algorithme offrant une estimation dite haute** de la population a une sensibilité de 63,4% et une spécificité de 98,7%.

La spécificité est élevée mais il reste cependant de nombreux FP. Il est difficile d'interpréter la part de ces FP liée à une sous-déclaration dans Constances, ou à une identification via l'algorithme SNDS de patients non épileptiques. Il faut toutefois se rappeler que le nombre de patients identifiés comme épileptiques dans le SNDS est beaucoup plus important que le nombre attendu d'après la littérature (cf. paragraphe 5.1). En effet, comme discuté plus bas, l'algorithme inclut à tort des patients traités par un AE dans un autre contexte que l'épilepsie (par exemple : la prégabaline).

La sensibilité peut paraître basse et implique un nombre de FN élevé. Pourtant, nous avons sélectionné a priori tous les critères disponibles dans le SNDS permettant d'identifier les patients avec épilepsie, à savoir l'ALD, l'hospitalisation et la délivrance de médicaments AE.

L'analyse des FN nous a permis d'explorer pourquoi certains patients n'ont pas été identifiés dans l'algorithme SNDS et d'identifier d'éventuels nouveaux critères de sélection (cf. paragraphe 5.2.3).

**Les estimations par rapport à l'algorithme offrant une estimation dite basse** de la population indiquent une sensibilité à 44,3% et une spécificité de 99,6%.

Cette spécificité très élevée est attendue, car les critères d'inclusion sont eux même très spécifiques : ALD pour épilepsie grave, ou hospitalisation pour épilepsie ou médicament AE dit « spécifique » de l'épilepsie. Par conséquent, le nombre de FP apparait moindre que pour l'algorithme en population haute. De la même manière, cette sensibilité, plus faible que l'algorithme en population haute, est attendue. En effet, la population est dite basse car elle exclut de nombreux patients épileptiques traités par un AE non spécifique de l'épilepsie. Or, de nombreux médicaments AE sont indiqués non seulement dans la prise en charge de l'épilepsie mais également dans la prise en charge d'autres pathologies, et en particulier des médicaments indiqués en première intention pour certains syndromes épileptiques (tels que la lamotrigine ou l'acide valproïque). Par conséquent cet algorithme (en population basse) est moins sensible (que celui en population haute) puisqu'il ne prend pas en compte ces médicaments.

A noter que la valeur plutôt basse de la sensibilité observée pour les deux algorithmes peut aussi être liée à la présence de patients avec CNEP ayant déclaré une épilepsie. Ces patients ne relevant pas d'une prise en charge de l'épilepsie telle que définie dans l'algorithme SNDS, s'ils ont déclaré une épilepsie dans les questionnaires Constances, ils contribuent à surestimer le nombre de FN. Il en est de même pour les diagnostics d'épilepsie posés à tort.

**Au total,** la précision des algorithmes n'est pas très élevée (respectivement 17,8% et 35% en population haute et basse), en comparaison à l'algorithme « contrôle » (association ALD/Hospitalisation) (58,1%) (cf. Tableau 7). Ces algorithmes en population haute et basse permettent toutefois de gagner en sensibilité dans l'identification de patients avec épilepsie par rapport à l'algorithme « contrôle » (+33 points et +14 points respectivement), sans trop perdre en spécificité, notamment pour l'algorithme en population basse (-0,3 points versus -1,2 points en population haute). **Rappelons que les caractéristiques estimées pour valider les algorithmes SNDS grâce à la cohorte CONSTANCES doivent être interprétées avec précaution, notamment en raison de la sous-déclaration.**

*6.2.3   Critères d'inclusion/exclusion*

L'analyse des critères d'inclusion/exclusion ci-après permet d'étudier les différentes possibilités d'amélioration des algorithmes.

**L'ALD pour épilepsie grave et l'hospitalisation pour épilepsie sont des critères fiables, directement liés à un diagnostic d'épilepsie**. Ce sont d'ailleurs les critères choisis par l'assurance maladie pour la cartographie des pathologies et des dépenses associées à l'épilepsie. **La délivrance de médicaments AE n'est pas nécessairement liée à un diagnostic d'épilepsie. La fiabilité de ce critère est donc moins certaine que l'ALD ou l'hospitalisation**. La déclaration d'une épilepsie dans Constances concerne respectivement 19% et 11% des patients traités par AE (sans ALD ni hospitalisation) dans les populations dites « basse » et « haute ». Ce taux de déclaration (plus faible que pour l'ALD et l'hospitalisation) peut être la combinaison de deux facteurs : une déclaration moins fréquente chez ces patients moins sévères comme dit plus haut, ou **l'identification à tort de patients non épileptiques.** Pour étudier ce dernier facteur, nous avons analysé plus particulièrement les FP.

L'analyse des FP suggère de **modifier le critère « délivrance de médicament AE » de deux manières.**

Premièrement, le nombre de FP important peut être dû notamment à **l'inclusion de patients n'ayant qu'une seule délivrance d'AE**. En effet, **imposer la présence d'au moins deux délivrances à l'algorithme en population haute améliore sa spécificité (99,3% versus 98,7%) et par conséquent sa précision (28,6% versus 17,8%)**, en baissant très peu sa sensibilité (cf. Annexe 8.6). De même, **imposer ce critère à l'algorithme en population basse permet d'améliorer légèrement sa spécificité (99,7% versus 99,6%) et par conséquent sa précision (41,8% versus 35%)** (cf. Annexe 8.68.7). Au total, la modification du critère « nombre délivrance de médicament AE » des algorithmes, en imposant au moins 2 délivrances à au moins 30 jours d'intervalle, permet d'augmenter leur spécificité et donc leur précision (+10 points et +7 points respectivement en population haute et basse), sans trop modifier la sensibilité (-0,8 points) (cf. Tableau 7).

Deuxièmement, nous savons que nos algorithmes SNDS peuvent **inclure à tort des patients traités par un AE ayant une autre indication que l'épilepsie**. Même s'il existe une sous-déclaration des patients avec épilepsie dans le(s) questionnaire(s) d'inclusion et/ou de suivi pouvant expliquer le nombre de FP élevé, et même si nous avons essayé d'exclure les patients avec un autre contexte que l'épilepsie dans la population haute, **il subsiste des patients traités par AE pour une autre indication que l'épilepsie sélectionnés par notre algorithme SNDS** (estimation haute surtout, mais également estimation basse). **L'algorithme en population haute semble surtout inclure à tort des patients traités par prégabaline et gabapentine**. Ces médicaments de la classe des AE ont comme autres indications que l'épilepsie la douleur neuropathique et l'anxiété généralisée, et semblent être délivrés majoritairement dans ces indications, d'après les données de la littérature (Derry et al., 2019; Greenblatt and Greenblatt, 2018). **Supprimer ces deux AE (prégabaline et gabapentine) des critères d'inclusion permet d'améliorer la précision de l'algorithme en population haute (45,3%)** (cf. Annexe 0). **L'algorithme en population basse semble quant à lui inclure à tort des patients traités par clonazépam**. Pourtant, ce médicament n'a pas d'autres indications AMM que l'épilepsie. Toutefois, en 2011, des nouvelles conditions de prescriptions et de délivrances de ce médicament ont été mises en place par l'ANSM (anciennement AFSAPPS) pour limiter la forte proportion de prescriptions hors-AMM pour le traitement de la douleur, de l'anxiété, des troubles du sommeil ou d'autres troubles psychiatriques: la prescription par des psychiatres en remplacement du Rohypnol® (flunitrazépam) pour le traitement de l'insomnie ; un potentiel d'abus, de dépendance et d'usage détourné chez les toxicomanes; l'utilisation du médicament dans la soumission chimique (l'administration à l'insu d'une personne).[9] **Supprimer le clonaézpam de l'algorithme en population basse permet d'améliorer quelque peu ses caractéristiques** (cf. Annexe 8.9).

Pour éviter d'inclure à tort des patients non épileptiques dans l'algorithme en population haute, nous avons appliqué des critères d'exclusion sur les délivrances de médicaments AE. Ces critères ne semblent pas suffisants mais ils permettent toutefois d'exclure un grand nombre de patients a priori non épileptiques. En effet, l'algorithme SNDS identifie 4234 patients avec une délivrance de médicament AE, sans ALD, ni hospitalisation, dont 1933 (46%) sont exclus car ils présentent un autre contexte pathologique, et parmi eux seulement 45 patients sont des FN. Ceci suggère que l'**algorithme exclut à tort seulement 2% de patients** (45/1933).

Notons enfin que certains patients n'ont pas pu être inclus avec l'algorithme SNDS parce qu'ils présentaient des critères d'inclusion en dehors de la période étudiée (avant ou après l'année 2018**). Il aurait fallu étendre la période d'analyse des données du SNDS pour l'inclusion des patients sur plusieurs années plutôt qu'une seule**. Nous n'avons pas pu le réaliser dans cette étude en raison de la mise à disposition des données et des contraintes calendaires. D'autres critères d'inclusion pourraient être envisagés, tels que les hospitalisations en soins de suite et de réadaptation (SSR) ou encore l'EEG.

---

[9] Plusieurs références sur l'usage détourné du RIVOTRIL® (clonazépam) dans la littérature. Citons ce rapport de l'EHESP : https://documentation.ehesp.fr/memoires/2011/phisp/cardon.pdf

*Tableau 7 : Comparaison des caractéristiques des algorithmes SNDS.*

| | Population Haute | Population Haute ≥2 délivrances à 30 jours d'intervalle | Population Haute Sans prégabaline et gabapentine | Population basse | Population basse ≥2 délivrances à 30 jours d'intervalle | Population basse Sans clonazépam | Population « contrôle » |
|---|---|---|---|---|---|---|---|
| Sensibilité | 63,4% | 62,6% | 62,1% | 44,3% | 43,5% | 44% | 30,3% |
| Spécificité | 98,7% | 99,3% | 99,7% | 99,6% | 99,7% | 99,8% | 99,9% |
| VPP | 17,8% | 28,6% | 45,3% | 35% | 41,8% | 43,3% | 58,1% |
| VPN | 99,8% | 99,8% | 99,8% | 99,8% | 99,8% | 99,8% | 99,7% |

# 7  Conclusion

Ce travail a permis d'apporter des éléments sur la validation des algorithmes SNDS proposés par la mission data pour étudier la pratique de prise en charge des patients avec épilepsie. Il est difficile de conclure sur les résultats obtenus en raison des limites associées à la déclaration de l'épilepsie dans Constances, avec à la fois des éventuelles déclarations à tort d'une épilepsie (erreur de diagnostic, CNEP), et surtout des sous-déclarations dont l'ordre de grandeur est estimé à au moins 1 pour 3. Une dernière limite importante est celle associée à la non-représentativité des patients avec épilepsie dans la cohorte Constances par rapport à la population générale (prévalence deux fois moins importante dans Constances, et peut être patients avec une épilepsie plus sévère).

Malgré ces limites, ces résultats nous confirment que si l'algorithme en population basse a une spécificité élevée, il n'est pas suffisamment sensible pour détecter l'ensemble de la population épileptique. Concernant l'algorithme en population haute, les résultats indiquent qu'il a une sensibilité plus élevée que l'algorithme en population basse, grâce au critère d'inclusion de l'ensemble des AE. Ce critère entraine toutefois l'inclusion à tort de patients non épileptiques et la précision de cet algorithme reste très basse, et inférieure à celle estimée pour l'algorithme en population basse.

Les deux algorithmes incluent à tort des patients non épileptiques par la présence de délivrance de médicaments AE dans un autre contexte que l'épilepsie, et sont donc moins précis qu'un algorithme « contrôle » qui ne comprendrait pas de critère de délivrance de médicaments AE mais uniquement la présence d'une ALD ou d'une hospitalisation pour épilepsie. Ces algorithmes en population haute et basse permettent toutefois de gagner en sensibilité dans l'identification de patients avec épilepsie par rapport à cet algorithme « contrôle » (+33 points et +14 points respectivement), sans trop perdre en spécificité, notamment pour l'algorithme en population basse (-0,3 points versus -1,2 points en population haute).

Pour pallier au manque de précision dû à l'ajout de ce critère de délivrance de médicaments antiépileptiques, une solution consiste notamment à affiner ce critère en proposant au moins deux délivrances plutôt qu'une seule. Ceci permet d'améliorer la précision des algorithmes (notamment pour l'algorithme en population haute) sans altérer (ou peu) leur sensibilité. De plus, certains médicaments AE pourraient être exclus des algorithmes car a priori très largement utilisés dans une autre indication, par exemple : la prégabaline et la gabapentine. Les supprimer des critères de sélection permet d'améliorer nettement la précision de l'algorithme en population haute (VPP=45,3%), pour se rapprocher de la précision de l'algorithme contrôle évoqué ci-dessus (VPP=58,1%), tout en augmentant la sensibilité. Cet algorithme devient alors plus performant que l'algorithme en population basse, et ce grâce à l'application de critères d'exclusion sur les AE.

Il faut évidement avoir en tête que la précision « vraie » des algorithmes reste supérieure à ces estimations, en raison des limites évoquées en introduction à propos de la cohorte Constances (sous-déclaration et non représentativité), sans que nous puissions la déterminer précisément.

Pour étudier les caractéristiques réelles de nos algorithmes, il faudrait utiliser une autre source de données plus représentative de la population française de patients avec épilepsie, incluant notamment la population pédiatrique. A notre connaissance, une telle base n'existe pas.

Soulignons aussi que nous n'avons pas identifié dans cette étude de nouveaux critères parmi les ALD et les hospitalisations permettant d'identifier les patients avec épilepsie. Toutefois, notre recherche s'est limitée au champ MCO. L'inclusion des séjours hospitaliers pour épilepsie en établissement SSR peut être une piste d'amélioration des algorithmes. Une autre piste d'amélioration serait simplement d'étendre la période d'analyse des données du SNDS pour l'inclusion des patients à plusieurs années. L'EEG pourrait aussi être un évènement traceur à prendre en compte comme critère d'inclusion.

# 8 Annexes

## 8.1 Annexe 1 : Liste des médicaments antiépileptiques ; les médicaments spécifiques de l'épilepsie (n'ayant comme indication AMM que l'épilepsie) apparaissent en gras.

| Substance active | Classe ATC |
|---|---|
| Acide valproïque | N03AG01 |
| **Brivaracétam** | N03AX23 |
| **Cannabidiol** | N03AX24 |
| Carbamazépine | N03AF01 |
| **Cénobamate** | N03AX25 |
| Clobazam | N05BA09 |
| **Clonazépam** | N03AE01 |
| Diazépam (Gel rectal ou Susp. Inj.) | N05BA01 |
| **Eslicarbazépine** | N03AF04 |
| **Ethosuximide** | N03AD01 |
| **Felbamate** | N03AX10 |
| **Fenfluramine** | N03AX26 |
| **Fosphénytoïne** | N03AB05 |
| Gabapentine | N03AX12 |
| **Lacosamide** | N03AX18 |
| Lamotrigine | N03AX09 |
| **Lévétiracétam** | N03AX14 |
| **Mesuximide** | N03AD03 |
| **Midazolam (BUCCOLAM®)** | N05CD08 |
| **Oxcarbazépine** | N03AF02 |
| **Perampanel** | N03AX22 |
| **Phénéturide** | N03AX12 |
| **Phénobarbital** | N03AA02 |
| **Phénytoïne** | N03AB02 |
| Prégabaline | N03AX16 |
| **Primidone** | N03AA03 |
| **Rufinamide** | N03AF03 |
| **Stiripentol** | N03AX17 |

| **Sultiame** | N03AX03 |
| **Tiagabine** | N03AG06 |
| Topiramate | N03AX11 |
| **Vigabatrin** | N03AG04 |
| **Zonisamide** | N03AX15 |

8.2 Annexe 2 : Libellés des réponses aux questionnaires ayant conduit à classer le patient en « suspicion d'épilepsie »

| Libellé | FN |
|---|---|
| Crise probable d'épilepsie | Oui |
| 1 crise épilepsie | Oui |
| Suspicion de crise d'épilepsie | |
| Epilepsie crise isolée | Oui |
| Anomalies épileptiques consécutives à 1 unique crise en 2007 | |
| Début crise d'épilepsie | |

8.3   Annexe 3 : Nombre de patients (et %) parmi les VP et parmi les FP, selon les critères de sélection de l'algorithme SNDS en population haute.

| Vrais positifs (N=437) | | |
|---|---|---|
| | N | % |
| AE ONLY | 228* | 52,17 |
| AE+ALD | 110 | 25,17 |
| AE+ALD+HOSPIT | 18 | 4,12 |
| AE+HOSPIT | 16 | 3,66 |
| ALD ONLY | 45 | 10,3 |
| ALD+HOSPIT | 12 | 2,75 |
| HOSPIT ONLY | 8 | 1,83 |
| Faux Positifs (N= 2 016) | | |
| | N | % |
| AE ONLY | 1865** | 92,51 |
| AE+ALD | 33 | 1,64 |
| AE+ALD+HOSPIT | 6 | 0,3 |
| AE+HOSPIT | 25 | 1,24 |
| ALD ONLY | 36 | 1,79 |
| ALD+HOSPIT | 3 | 0,15 |
| HOSPIT ONLY | 48 | 2,38 |

*dont 222 patients avec au moins deux délivrances à au moins 30 jours d'intervalle

**dont 925 patients avec au moins deux délivrances à au moins 30 jours d'intervalle

8.4 Annexe 4 : Nombre (et pourcentage) de patients identifiés par une délivrance de médicaments AE, sans ALD ni hospitalisation selon la mention d'épilepsie dans CONSTANCES (groupe FP ou VP).

| | Algorithme en population haute | | | | Algorithme en population basse | | | |
|---|---|---|---|---|---|---|---|---|
| | N | FP | % par rapport à N | % par rapport à FP | N | FP | % par rapport à N | % par rapport à FP |
| **Patients distincts avec AE sans ALD ni hospitalisation pour épilepsie** | 2093 | 1865 | | | 511 | 415 | | |
| CARBAMAZEPINE | 127 | 96 | 75,6 | 5,1 | 0 | | | |
| CLOBAZAM | 73 | 60 | 82,2 | 3,2 | 0 | | | |
| CLONAZEPAM | 19 | 19 | 100 | 1 | 172 | 170 | 98,8 | 41 |
| DIAZEPAM | 16 | 16 | 100 | 0,9 | 50 | 50 | 100 | 12 |
| ESLICARBAZEPINE | 5 | 3 | 60,0 | 0,2 | 10 | 6 | 60,0 | 1,4 |
| ETHOSUXIMIDE | 3 | 3 | 100 | 0,2 | 3 | 3 | 100 | 0,7 |
| GABAPENTINE | 346 | 341 | 98,6 | 18,3 | 0 | | | |
| LACOSAMIDE | 15 | 9 | 60,0 | 0,5 | 19 | 11 | 57,9 | 2,7 |
| LAMOTRIGINE | 126 | 52 | 41,3 | 2,8 | 0 | | | |
| LEVETIRACETAM | 107 | 60 | 56,1 | 3,2 | 143 | 88 | 61,5 | 21,2 |
| MIDAZOLAM | 2 | 2 | 100 | 0,1 | 2 | 2 | 100,0 | 0,5 |
| OXCARBAZEPINE | 23 | 15 | 65,2 | 0,8 | 71 | 59 | 83,1 | 14,2 |
| PERAMPANEL | 3 | 2 | 66,7 | 0,1 | 4 | 3 | 75,0 | 0,7 |
| PHENOBARBITAL | 27 | 8 | 29,6 | 0,4 | 33 | 13 | 39,4 | 3,1 |
| PHENYTOINE | 4 | 3 | 75,0 | 0,2 | 0 | | | |
| PREGABALINE | 1202 | 1198 | 99,7 | 64,2 | 0 | | | |
| PRIMIDONE | 20 | 20 | 100 | 1,1 | 25 | 25 | 100,0 | 6 |
| TOPIRAMATE | 58 | 53 | 91,4 | 2,8 | 0 | | | |
| VALPROIQUE ACIDE | 65 | 18 | 27,7 | 1 | 0 | | | |
| ZONISAMIDE | 4 | 2 | 50,0 | 0,1 | 8 | 4 | 50,0 | 1 |

8.5 Annexe 5 : Nombre de patients (et %) parmi les VP et parmi les FP, selon les critères de sélection de l'algorithme SNDS en population basse.

| Vrais positifs (N=305) | | |
| --- | --- | --- |
| | N | % |
| AE ONLY | 96* | 31,48 |
| AE+ALD | 74 | 24,26 |
| AE+ALD+HOSPIT | 19 | 6,23 |
| AE+HOSPIT | 12 | 3,93 |
| ALD ONLY | 81 | 26,56 |
| ALD+HOSPIT | 11 | 3,61 |
| HOSPIT ONLY | 12 | 3,93 |
| Faux positifs (N=566) | | |
| | N | % |
| AE ONLY | 415** | 73,32 |
| AE+ALD | 29 | 5,12 |
| AE+ALD+HOSPIT | 5 | 0,88 |
| AE+HOSPIT | 29 | 5,12 |
| ALD ONLY | 40 | 7,07 |
| ALD+HOSPIT | 4 | 0,71 |
| HOSPIT ONLY | 44 | 7,77 |

*dont 91 patients avec au moins deux délivrances à au moins 30 jours d'intervalle

**dont 267 patients avec au moins deux délivrances à au moins 30 jours d'intervalle

8.6    Annexe 6 : Tableau croisé des patients dans le SNDS et des patients dans CONSTANCES avec épilepsie – Algorithme en population haute utilisant au moins 2 délivrances à ≥30 jours d'intervalle pour le critère médicament.

|  | CONSTANCES + | CONSTANCES - | Total |
|---|---|---|---|
| **SNDS +** | 431 = VP | 1 076 = FP | **1 507** |
| **SNDS -** | 258 = FN | 155 054 = VN | 155 312 |
| Total | **689** | 156 130 | 156 819 |

La sensibilité est égale à : $\frac{VP}{VP+FN} = \frac{431}{689} = \mathbf{63}\%$

La spécificité est égale à : $\frac{VN}{VN+FP} = \frac{155\,054}{156\,130} = \mathbf{99,3}\%$

La VPP est égale à : $\frac{VP}{VP+FP} = \frac{431}{1507} = \mathbf{28,6}\%$

La VPN est égale à : $\frac{VN}{VN+FN} = \frac{155\,054}{156\,819} = \mathbf{99,8}\%$

8.7    Annexe 7 : Tableau croisé des patients dans le SNDS et des patients dans CONSTANCES avec épilepsie – Algorithme en population basse utilisant au moins 2 délivrances à ≥30 jours d'intervalle pour le critère médicament.

|  | CONSTANCES + | CONSTANCES - | Total |
|---|---|---|---|
| **SNDS +** | 300 = VP | 418 = FP | **718** |
| **SNDS -** | 389 = FN | 155 712 = VN | 156 101 |
| Total | **689** | 156 130 | 156 819 |

La sensibilité est égale à : $\frac{VP}{VP+FN} = \frac{300}{689} = \mathbf{43,5}\%$

La spécificité est égale à : $\frac{VN}{VN+FP} = \frac{155\,712}{156\,130} = \mathbf{99,7}\%$

La VPP est égale à : $\frac{VP}{VP+FP} = \frac{300}{689} = \mathbf{41,8}\%$

La VPN est égale à : $\frac{VN}{VN+FN} = \frac{155\,712}{156\,101} = \mathbf{99,8}\%$

8.8 Annexe 8 : Tableau croisé des patients dans le SNDS et des patients dans CONSTANCES avec épilepsie – Algorithme en population haute en supprimant la prégabaline et la gabapentine des critères d'inclusion.

|  | CONSTANCES + | CONSTANCES - | Total |
|---|---|---|---|
| **SNDS +** | 428 = VP | 516 = FP | **944** |
| **SNDS -** | 261 = FN | 155 614 = VN | 155 875 |
| Total | **689** | 156 130 | 156 819 |

La sensibilité est égale à : $\frac{VP}{VP+FN} = \frac{428}{689} = \mathbf{62,1}\%$

La spécificité est égale à : $\frac{VN}{VN+FP} = \frac{155\,614}{156\,130} = \mathbf{99,7}\%$

La VPP est égale à : $\frac{VP}{VP+FP} = \frac{428}{944} = \mathbf{45,3}\%$

La VPN est égale à : $\frac{VN}{VN+FN} = \frac{155\,614}{155\,875} = \mathbf{99,8}\%$

8.9 Annexe 9 : Tableau croisé des patients dans le SNDS et des patients dans CONSTANCES avec épilepsie – Algorithme en population basse en supprimant le clonazépam des critères d'inclusion.

|  | CONSTANCES + | CONSTANCES - | Total |
|---|---|---|---|
| **SNDS +** | 303 = VP | 396 = FP | **699** |
| **SNDS -** | 386 = FN | 155 734 = VN | 156 120 |
| Total | **689** | 156 130 | 156 819 |

La sensibilité est égale à : $\frac{VP}{VP+FN} = \frac{303}{689} = \mathbf{44}\%$

La spécificité est égale à : $\frac{VN}{VN+FP} = \frac{155\,734}{156\,130} = \mathbf{99,7}\%$

La VPP est égale à : $\frac{VP}{VP+FP} = \frac{303}{689} = \mathbf{43,3}\%$

La VPN est égale à : $\frac{VN}{VN+FN} = \frac{155\,734}{156\,120} = \mathbf{99,8}\%$

# 9 Abréviations

| AE | antiépileptique |
|---|---|
| ALD | affection longue durée |
| AMM | autorisation de mise sur le marché |
| ATC | anatomique thérapeutique et chimique |
| ATU | autorisation temporaire d'utilisation |
| CES | centres d'examens de santé |
| CIM-10 | classification internationale des maladies 10ème version |
| CIP | code identifiant de présentation |
| CNAM | Caisse nationale d'assurance maladie |
| CNEP | crise non épileptique psychogène |
| DAS | diagnostic associé |
| DCIR | datamart de consommations inter-régimes |
| DP | diagnostic principal |
| DR | diagnostic relié |
| EEG | électroencéphalogramme |
| FN | faux négatif |
| FP | faux positif |
| GHS | Groupe homogène de séjour |
| HAS | Haute autorité de santé |
| MCO | médecine chirurgie obstétrique |
| PMSI | programme de médicalisation du système d'information |
| SNDS | système national des données de santé |
| SSR | Soins de suite et de réadaptation |
| STSS | stratégie de transformation du système de santé |
| VN | vrai négatif |
| VP | vrai positif |

| VPN | valeur prédictive négative |
| VPP | valeur prédictive positive |

## 10 Accès aux données et réglementation

Dans le cadre du respect du règlement général sur la protection des données, les données de ce projet seront mises à disposition par le centre d'accès sécurisé aux données via une bulle sécurisée. Le projet a fait l'objet d'une déclaration à la Commission nationale informatique et liberté selon la référence MR004 et d'une analyse d'impact relative à la protection des données. Ce protocole a été soumis au comité scientifique et éthique de la cohorte Constances.

## 11 Communication

Ce rapport a fait l'objet d'une communication orale lors du congrès EMOIS qui s'est déroulé à Lille du 4 au 5 avril 2024 : Degremont, A., Bisquay, C., Jachiet, P-A. J Epidemiol Popul Health. doi : 10.1016/j.jeph.2024.202232.

# 12 Bibliographie